\documentclass[12pt]{article}

\usepackage{amssymb,amsmath,bbm,tensor,amsthm} 
\usepackage{lmodern, microtype} 
\usepackage[dvipsnames]{xcolor} 
\usepackage{setspace} 
\usepackage{hyperref} 
\usepackage[left= 1in,right=1in,top=1in,bottom=1in]{geometry} 
\usepackage{comment} 
\usepackage{authblk} 
\usepackage{graphicx} 
\usepackage{caption} 
\usepackage{subcaption}
\usepackage[shortlabels]{enumitem} 
\usepackage{epigraph} 
\usepackage[utf8]{inputenc}
\usepackage{csquotes}

\usepackage{doi} 
\usepackage[natbibapa]{apacite} 

\hypersetup{
    colorlinks=true, 
    linkcolor=Blue,
    citecolor=OliveGreen,
    urlcolor=MidnightBlue,
    filecolor=magenta,
    anchorcolor=black
}

\let\oldmarginpar\marginpar
\renewcommand\marginpar[1]{\oldmarginpar{\color{red}\raggedright\scriptsize #1}}
\newcommand{\Sec}[1] {Section~#1}
\newcommand{\pb}[2]{\ensuremath{\lf\{#1,#2 \rt\}}}

\newcommand{\deby}[2]{\ensuremath{\frac{\rm d #1}{\rm d #2}}}
\newcommand{\diby}[2]{\ensuremath{\frac{\partial #1}{\partial #2}}}

\def\lf {\ensuremath{\left}} 
\def\rt {\ensuremath{\right}}
\def\de {{\rm d}} 

\def\Sec {Section~}

\title{An account of the arrow of time when scale is surplus}

\author[1,2]{{\bf Sean Gryb}\thanks{email: \href{mailto:s.b.gryb@rug.nl}{s.b.gryb@rug.nl}, website: \href{https://seangryb.wordpress.com}{seangryb.wordpress.com}}}
\author[1]{ {\bf Simon Friederich}\thanks{email: \href{mailto:s.m.friederich@rug.nl}{s.m.friederich@rug.nl}} }
\affil[1]{ {\it University College Groningen}, University of Groningen }
\affil[2]{{\it Faculty of Philosophy}, University of Groningen}

\date{\today}

\begin{document}

\maketitle

\begin{abstract}
    Existing accounts of the cosmological arrow of time face a dilemma: generalist approaches that posit time-asymmetric laws lack independent motivation, while particularist approaches that invoke a Past Hypothesis face serious conceptual and explanatory problems. We propose a novel account that dispenses with the need for both time-asymmetric laws and a Past Hypothesis. Instead, it is centred around a symmetry argument that reveals attractors and so-called \emph{Janus points}. The main idea is that the global spatial scale of the Universe is not empirically accessible, and should therefore be treated as surplus structure. By contrast, the Hubble parameter, which encodes the relative rate of change of scale, \emph{is} empirically accessible. Once the scale redundancy is eliminated, the empirically meaningful description of cosmic dynamics involves a drag-like variable that transforms in a way that preserves time-reversal invariance. The resulting space of dynamical possibilities possesses universal attractors and Janus points, giving rise to particular states in which observers experience a cosmological arrow of time akin to our own. We illustrate the proposal by applying it to cosmological and gravitational $N$-body models, showing how it accounts, respectively, for the rapid cooling of the early universe and its relative smoothness.
\end{abstract}

\clearpage

\tableofcontents

\clearpage

\section{Introduction}\label{sec:intro}

\subsection{Motivation}

``Arrow of Time'' (AoT) is a convenient label for a persistent and widespread tendency of processes in our universe to display a preferred temporal orientation. We see cups shattering into shards, but not shards reforming into cups. This tendency can often be quantified in terms of monotonic gradients by counting, for example, the density of grey hairs on one's head or by computing the entropy of a system. For the purposes of this paper, we will take the AoT to be an asymmetry \emph{in}, but not necessary \emph{of}, time in which physical processes express a numerically large temporal gradient in at least one of their features.

While the AoT pervades our lived experience of the macroscopic world, its origin is a puzzle from the perspective of the microscopic laws of fundamental physics, which are conspicuously devoid of any structure that could explain the sheer amount of time asymmetry seen in the world. The problem of explaining the origin of the AoT despite the near-time asymmetry of the known fundamental laws is known as the \emph{problem of the AoT}.

Existing explanations of the AoT tend to posit some sort of constraint on the dynamical possibilities of a theory. One option is to propose laws that break time-reversal invariance enough to explain the AoT, but in a way that is numerically compatible with observations. This could involve, for example, postulating an explicit collapse model of quantum mechanics that would introduce a significant time asymmetry without diverging substantively from the predictions of quantum physics. Another, more popular, option is to introduce a constraint, usually called a \emph{Past Hypothesis}, on the state of the early Universe. The idea is to explain the experienced AoT by hypothesising that the Universe started in a very special state, and that the natural dynamical evolution of this state is towards more typical states like the ones experienced today. While these explanations have been the subject of immense philosophical discussion since the advent of statistical mechanics, both options currently face serious objections related to basic explanatory limitations and difficulties arising in concrete implementations.\footnote{We will describe these two explanatory approaches more carefully in \Sec\ref{sec:Price taxonomy} and detail their main objections in \Sec\ref{sec:the_dilemma}.}

In this paper, we will first sharpen this dilemma for all extant solutions to the problem of the AoT and then propose a new kind of solution. Our solution neither invokes a Past Hypothesis nor breaks time-reversal invariance. Instead, we define a new scenario for explaining the AoT that begins with a symmetry argument and ends with the local emergence of a temporal arrow. In our scenario, there is no AoT at the level of the definition of the theory alone. Rather, an AoT is experienced once there are observers existing in states that obey particular dynamical conditions.

Our analysis is based on a symmetry argument in which the scale factor of Universe is shown to \emph{not} be necessary for explaining the relevant empirical phenomena. In contrast, the Hubble parameter, which quantifies the relative rate of change of the scale factor, \emph{is} found to be necessary. We therefore treat the scale factor, but not the Hubble parameter, as a gauge degree of freedom of the theory.

The resulting framework leads to a new understanding of cosmological law. In the new picture, it is simultaneously possible for laws to have a friction-type term while \emph{still} being time-reversal invariant. In this setting, we identify a dynamical variable that plays the role of a \emph{drag} coefficient in the way that it interacts with the remaining degrees of freedom of the system, but that transforms in such a way that the total system retains time-reversal invariant. For dynamical systems of this kind, we show that the solutions of the theory can start and end in attractors, and that drag-free points, which we identify with \emph{Janus points} following a terminology introduced by Julian Barbour,\footnote{See, for example, \cite{barbour2013gravitational, Barbour:2014bga}, where the terminology was introduced and \cite{barbour2020janus}, where further concepts were developed.} serve like midpoints in the dynamical evolution.

This scenario, which we call a \emph{Janus--Attractor scenario}, provides a general explanation of the AoT when realised. This is because observers in states close to an attractor see an AoT pointing from the Janus point to the attractor. In cases where the degree of the AoT can be quantified, the notion of closeness to the attractor defines a monotonic quantity, allowing for an explicit model of the numerical gradients in question. The overall picture, detailed in \Sec\ref{sec:JA scenario}, is summarised in Figure~\ref{fig:JA scenario}.

We then proceed to show that our general scenario is realised in two concrete models that, we argue, explain important aspects of the AoT in our actual Universe. The first model is a cosmological Friedmann--Lema\^itre--Robertson--Walker (FLRW) model of the Universe suitable for understanding large-scale, global evolution. We show that, under natural physical assumptions, this model can explain key aspects of the cosmological AoT seen in our Universe through the behaviour of the Hubble parameter. The second model is a Newtonian $N$-body gravitational model suitable for understanding the local evolution of galaxies and clusters of galaxies in our Universe. We show that, when global scale is treated as a gauge degree of freedom in these models, early states are typically smooth --- contrary to standard intuitions. This non-standard result is possible because the natural measures on scale-free systems are not time-independent and, therefore, the natural notions of typicality depend on the kinds of states the system finds itself in at any given time. Together, these two models explain many of the puzzling features of the AoT.

We conclude our analysis by showing how our new account, based on attractors and Janus points, evades the known problems faced by existing accounts of the arrow of time.

\subsection{Roadmap}

In \Sec\ref{sec:Price taxonomy}, we review standard accounts of the AoT and, in \Sec\ref{sec:the_dilemma} we detail a number of objections raised against each. In \Sec\ref{sec:DS as gauge}, we motivate the symmetry argument at the basis of our proposal and sketch the general properties of the theory invariant under the relevant surplus structure. This leads to our general account of the AoT in terms of a Janus-Attractor scenario, which we describe in \Sec\ref{sec:general JA}. In \Sec\ref{sec:explanatory target}, we narrow down our explanatory target by focusing on two empirical problems that, we argue, account for many aspects of our experienced AoT. We then show how to use our framework to solve these problems in the context of two concrete models. These models are given in \Sec\ref{sec:JA and AoT}. Finally, in \Sec\ref{sec:reconsidering objections}, we show how our approach avoids the objections raised against standard accounts.

The results presented here extend and synthesise the analysis developed in \cite{gryb:phil_thesis}.

\section{Price's taxonomy}\label{sec:Price taxonomy}

\cite{price2002boltzmann} provides a helpful taxonomy for distinguishing different existing approaches to explaining the AoT. Price distinguishes between ``causal-generalist'' and ``acausal-particularist'' accounts of the AoT. Causal-generalism is characterised as follows:
\begin{quote}
    On one side are what I shall call \textit{Causal-General} theories. These approaches take the explanandum to be, at least in part, a time-asymmetric \textit{generalisation}---the general fact that entropy never decreases, or some such. Broadly speaking---perhaps taking some liberties with the terms \textit{causal} and \textit{dynamical}---they seek a causal explanation of this general fact in dynamical terms. Approaches I take to fall under this heading include `interventionism' and certain appeals to asymmetric initial microscopic independence conditions, as well as to suggestions grounded on law-like asymmetries in the dynamical laws themselves. What unifies these diverse approaches, in my view, is their sense of the nature of the project. All of them seek a \textit{causal-explanatory} account of a \textit{time-asymmetric generalisation} about the physical world as we find it.  (p.\ 90)
\end{quote}
In contrast, according to acausal-particularism,
\begin{quote}
     All the time-asymmetry of observed thermodynamic phenomena resides in an existential or particular fact---roughly, the fact that physical processes in the known universe are constrained by a low entropy `boundary condition' in one temporal direction. Against the background of a \textit{time-symmetric} understanding of the normal behaviour of matter, this particular fact alone is sufficient to account for the observed asymmetry in thermodynamic phenomena. The task of explaining the observed asymmetry is thus the task of explaining a particular violation of contrast class (b)---a particular huge entropy gradient, in a world in which (roughly) none are to be expected. (p.\ 92)
\end{quote}
We adopt a variation of Price's terminology: Since for the purposes of this paper, we sidestep \emph{causal} questions about the AoT, we refrain from using the terminology of cause and effect. Instead, we focus on whether any given approach hypothesizes time-reversal invariant \textit{laws} or not.

In relation to laws, a complication arises from the fact that the laws of nature encoded in our best current theories are not completely time-reversal, or $T$-, invariant but only ``nearly'' $T$-invariant. In particular, the Standard Model of elementary particle physics is notably \emph{not} $T$-invariant. For example, the time-reversed dynamics of a ``left-handed'' electron are those of a ``right-handed'' anti-electron (positron). There is some disagreement among philosophers regarding the appropriate metaphysical implications of this lack of $T$-invariance. Some; e.g., \cite[Ch 7]{earman1989world}; argue that this supports substantivalism about temporal orientation while others; e.g., \cite{pooley2003handedness,price1997book}; disagree.\footnote{For a discussion of how such $T$-violations can be understood and a summary of some issues involved, see \cite{roberts2022reversing} --- particularly Chapter 7.}

But when it comes to the quantitative empirical question of whether the \emph{amount} of time asymmetry resulting from $T$-violations in the Standard Model is sufficient for providing an adequate explanation of the cosmological AoT, the overwhelming consensus is an emphatic `no'.\footnote{Though see \cite{salimkhani} for recent dissent.} In particular, $T$-violations in the Standard Model are neither able to explain the shear amount of smoothness seen at recombination nor the dramatic red-shift in the early Universe. Thus, while $T$-violation in the Standard Model does constitute evidence for time asymmetry in the known fundamental laws, this time asymmetry is not able to provide an adequate explanation for the AoT.

Given the numerical insignificance of the $T$-symmetry violations in the Standard Model, we regard our fundamental laws as $T$-invariant \emph{For All Practical Purposes (FAPP)}. With this terminology in place, we propose the following taxonomy of ``generalist'' and ``particularist'' accounts of the AoT:
\begin{itemize}
    \item[]{\bfseries Generalism:} accounts of the AoT according to which the true laws of nature are time-asymmetric and make the thermodynamic AoT an expected feature.
    \item[]{\bfseries Particularism:} accounts of the AoT according to which the laws are FAPP time-symmetric while there is some particular, contingent, fact that makes the AoT expected.
\end{itemize}

Examples of attempts to provide generalist accounts of the AoT involve using spontaneous collapse models of quantum mechanics to introduce fundamental and significant time asymmetry into the laws. A well-known example, advocated in \cite{albert2009time}, makes use of the spontaneous collapse theory introduced by \cite{ghirardi1986unified}.

Particularist theories, in contrast, all rely on some version of the Past Hypothesis (PH). The most common way to implement a PH is in terms of a low-entropy initial state. We will see some simple examples of this below. However, low entropy is not the only way to frame a PH. The \emph{Weyl curvature hypothesis}, discussed near the end of \Sec\ref{sec:global to local arrow}, is a PH that imposes a constraint on the spacetime geometry of the initial state, and such a constraint has no obvious (or proven) connection to any notion of entropy. Additionally, while smooth states in self-gravitating systems are certainly low-entropy, it's not clear whether the same can be said about states with the rapid cooling experienced in the early Universe, which led to entropy reservoirs that can be linked to the AoT.\footnote{We will explain why this is an explanatory concern in \Sec\ref{sec:explanatory target}.} Instead, what makes a PH is that the past state falls into a class of states (e.g. smooth, highly red-shifted or low in Weyl curvature) that are deemed to be atypical according to some reasonable measure on the state space of the theory. We will see the reason for this in a moment.

Price argues in favour of acausal-particularism or, in our terms, particularism. The basic idea behind this type of approach is based on an old argument by Boltzmann (see, for example, \cite{boltzmann1895certain}) regarding the thermodynamic AoT in free gases. In \cite{price2002boltzmann}, the approach is described as follows:
\begin{quote}
    [T]he basic character of Boltzmann's statistical approach is well known. Consider a system not currently in equilibrium, such as a vial of pressurised gas within a larger evacuated container. We want to know why the gas expands into the larger container when the vial is opened. We consider what possible future `histories' for the system are compatible with the initial set-up. The key to the statistical approach is the idea that, under a plausible way of counting possibilities, almost all the available microstates compatible with the given initial macrostate give rise to future trajectories in which the gas expands. It is possible---both physically possible, given the laws of mechanics, and epistemically possible, given what we know---that the actual microstate is one of the rare `abnormal' states such that the gas stays confined to the pressurised vial. But in view of the vast numerical imbalance between abnormal and normal states, the behaviour we actually observe is `typical', and therefore calls for no further explanation. There is no need for an asymmetric causal constraint to `force' the gas to leave the bottle---this is simply what we should expect it to do anyway, if our expectations about its initial state are guided by Boltzmann's probabilities.(p.\ 92)
\end{quote}
The assumption of the initial non-equilibrium state of the system, in this case the ``vial of pressurised gas within a larger evacuated container,'' is the key ingredient of the proposed explanation of the observed behaviour of the gas. Because the Boltzmann entropy is the log of the phase space volume of a macrostate, low-entropy states are atypical under the natural measure on phase space. If Boltzmannian entropy is low for some time $t_0$ (i.e., the state at $t_0$ is atypical), then entropy increase is expected for times $t>t_0$ (i.e., the states after $t>t_0$ are expected to become more typical). Thus, the riddle of the thermodynamic AoT is removed.  The only explanatory task left, the idea goes, is to account for why the state at the early time $t_0$ is so atypical.

It is key to the particularist argument that it is framed in terms of the \emph{typicality} of states: states of low-entropy, vanishing Weyl curvature and rapidly decreasing red-shift are all understood to be atypical. But once they are established, the particularist reasoning goes, the existence and pervasiveness of an AoT at later times is entirely expected.

Price defends particularism and opposes generalism. His main reason for doing so is that, as he sees it, generalists must \emph{also} explain, just like the particularists must, the highly unusual early state of the Universe. As a result, the overall generalist explanation of the AoT is less economical:
\begin{quote}
     It is surprising that such a stark contrast --- the invocation of one temporal asymmetry in the latter [Acausal-Particular] approach as against two in the former [Causal-General] --- seems to have received little explicit attention in the literature. The contrast suggests that \emph{prima facie}, at least, the Acausal-Particular approach has considerable theoretical advantage. To the extent that asymmetry is a theoretical `cost', the Causal-General approach is a great deal less economical than its rival. (p.\ 99)
\end{quote}

In other words, according to Price, the generalist is as burdened to explain the atypicality of the past state as the particularist. The bulk of Price's paper is devoted to an exorcism of the idea that a causal mechanism or ``engine'' would further contribute substantively to explaining \textit{why} entropy rises even after the assumption has been granted that it starts out low. Once the idea that such a mechanism is needed has been debunked, the motivation for generalism is undermined, as he sees it, and particularism is the most promising way forward.

We agree with Price that, inasmuch as generalists face the same challenge as particularists to account for why the very early Universe is in a highly atypical state, generalism provides no added benefits. But we disagree that the superiority of particularism is as clear as he takes it to be. In \Sec\ref{sec:the_dilemma}, we describe a dilemma according to which particularist and generalist accounts of the AoT both have highly unattractive features, seemingly leaving no good option available.

\section{The dilemma}
\label{sec:the_dilemma}

In this section, we outline a number of objections against, respectively, generalist and particularist approaches to explaining the AoT. Some of these objections are familiar from the philosophy of physics literature, but only some have been harnessed explicitly as objections against specific accounts of the AoT. We give the objections to generalism first, then the objections to particularism.

\subsection{Objections to generalism}

\begin{enumerate}[G.I]
    \item {\bfseries Objection from redundancy} This objection is just Price's primary criticism of generalism as already described in \Sec\ref{sec:Price taxonomy}. It says that any generalist approach must posit special initial conditions \textit{in addition to} time-asymmetric dynamics. This makes the move of postulating time-asymmetric dynamics redundant and undermines the rationale for generalism.
    \item {\bfseries Objection from lack of independent motivation:} The fundamental known laws of physics, as currently known and encoded in the Standard Model of elementary particle physics and general relativity, are time-symmetric in the FAPP sense defined above. These laws are extremely successful in describing the physics and microphysics underlying all known processes in nature, including those which instantiate the thermodynamic AoT on a macroscopic level. The empirical hints that we have towards new physical laws -- for instance the inability of the Standard Model of elementary particle physics to account for dark matter -- do not seem to be related to the time symmetry of the Standard Model and general relativity. There is thus no independent empirical motivation, beyond potentially the AoT itself, for believing that time-asymmetric laws would be correct instead. It is also unclear how any account that incorporates generalism could preserve the facts that are so well accounted for by time-symmetric laws while simultaneously accounting for the thermodynamic AoT in terms of its significant time-asymmetric aspects.
    \item {\bfseries Objection from historical progress} From a historical perspective, comparing the evolution of our understanding of the spatial dimensions with the time dimension, it would be surprising if the laws of nature turned out to be time-asymmetric after all. ``Naive'' physics, based on our everyday experience, may suggest that there is a principled difference not only between opposite time-directions ``past'' and ``future'', but also between opposite spatial directions ``up'' and ``down.'' But already in antiquity, when it was recognized, that the Earth is spherical, a more accurate scientific picture emerged that makes no principled distinction between any two opposite space direction . The distinction between ``up'' and ``down'' has since been understood in terms of a particular fact: the Earth, as the body that exerts the strongest gravitating pull on us, defines what counts locally as ``down.''

    Since Newtonian times, our best laws of physics --- from Newton's theory of gravitation to the Standard Model of elementary particle physics --- have treated the time-coordinate analogously to the space-coordinates in this respect. They do not treat the two time-directions qualitatively differently. It is very natural to believe that this feature is not accidental and that any more fundamental laws, to be discovered in the future, will treat the two time directions analogously ---  at least on the ``global'' level for which this is true for the spatial directions. The CPT-theorem of quantum field theory even mixes time-reversal with reversal of spatial coordinates and thus provides further support that this analogy is not accidental. It can of course not be ruled out that future physics might take a different course and treat the past and future directions radically differently. But this would seem to be an odd turn in the historical development of our understanding of time and its role in the laws of nature --- not something that we would otherwise expect. 
\end{enumerate}

These objections to generalism, in one form or another, have had a stinging impact in the philosophical community. While there are still attempts to conceive of explanations of the AoT by appealing to time-asymmetric laws, many modern approaches embrace particularism instead. But, as we will see next, particularism faces serious problems as well --- problems glossed over by Price in his endorsement.

\subsection{Objections to particularism}\label{sec:objs particularism}

Objections to particularism which in our view must be taken seriously include:

\begin{enumerate}[P.I]
    \item {\bfseries Objections from mathematical and conceptual ambiguity:} These objections arise from the fact that the particular state required by particularist explanations is difficult to characterise mathematically and motivate physically. The PH, as we have formulated it, requires that the early state be atypical. This requires a reasonable typicality measure for ``counting'' the states of a theory. At early enough times in the Universe's history, general relativistic degrees of freedom must also be taken into account. Unfortunately, recent research has brought to light several daunting difficulties involved in doing so.

    As discussed at length in \cite{earman2006past}, \cite{Wald:2012zf}, \cite{Corichi:2013kua} and \cite{Curiel:2015oea}; even in homogeneous cosmology, which is a dramatically simplified version of general relativity, there are serious ambiguities in defining a measure on the space of possible models of the theory. In \cite{Corichi:2010zp} and \cite{Ashtekar:2011rm}, it is noted that such ambiguities lead to estimates of the probability of inflation that differ by up to $85$ orders of magnitude (e.g., \cite{Kofman:2002cj} compared to \cite{Turok:2006pa}). Using more realistic models only adds to the problem. As is show in \cite{Wald:2012zf}, the infinite dimensional nature of the state spaces of perturbed cosmological models leads to considerable mathematical and interpretational problems.

    In addition to these arguments, \cite{ph_takedown} uses the same argument based on scaling symmetry that we will leverage to motivate our own proposal in \Sec\ref{sec:JA scenario} to show that for a PH to have explanatory force, it must make a distinction without a difference. This distinction arises by over-counting empirically equivalent states that are simply global rescalings of each other.\label{obj:ambiguity}

    \item {\bfseries Objections from the breakdown of thermodynamic assumptions:} These arise from a lack of precision often involved in defining \emph{the} entropy of the Universe or in connecting the typicality of the Universe's states to the notions of entropy relevant to the local thermodynamic arrow. Since the Universe is a complicated many-body self-gravitating systems, it is not clear whether there is a good notion of entropy that can be applied consistently to the Universe as a whole. Moreover, even if such a notion were available, the particularities of the dynamics of these systems raise questions about whether thermodynamic concepts are even appropriate to use at various levels.\label{obj:thermo_breakdown}
    \item {\bfseries Objections from lack of explanatory force:} These arise from the absence of any good candidate explanations for the particular fact used to explain the AoT itself. Without such an explanation, it is not clear whether particularist accounts can have any real explanatory force. Even \cite{price2002boltzmann} admits that a ``solution to this new puzzle [of explaining why the PH holds] is not yet in hand. Indeed, it is not yet clear what a solution would look like.'' (p.118) As pointed out by \cite{callender2004measures}, it is difficult to see what \textit{kind} of explanation there could even be for such a particular fact and whether demanding such an explanation is even warranted at all.\label{obj:explanation}
\end{enumerate}

\section{Reconsidering ``typical'' states of the Universe}\label{sec:DS as gauge}

The \textit{objection from lack of explanatory force} highlights that there is an inherent tension in the idea of making the AoT a \emph{typical, expected}, feature precisely by hypothesizing an \emph{atypical, unexpected}, early state. To make matters worse, the entire particularist framework is predicated on the existence of a natural measure on the state-space of theory that can be used to determine which types of Universe states are ``typical'' and which ones aren't. But the \emph{objections from mathematical and conceptual ambiguity} raise questions about whether there can ever be a clearly preferred choice of measure. In cosmological models, these ambiguities relate to the freedom to arbitrarily redefine the overall spatial scale of the Universe \citep{ph_takedown,Gryb:2021qix}. In what follows, we frame these observations in terms of a symmetry of the Universe called \emph{dynamical similarity} \citep{Sloan:2018lim} and argue that it should be considered a gauge symmetry of cosmology, i.e. a symmetry that connects physically identical cosmic histories (Dynamically Possible Models [DPMs], in what follows). We will use this concept to show that there is no natural time-independent measure on the space of cosmological DPMs. Instead, the natural gauge-invariant measures will be shown to be time-\emph{dependent} and, as such, unable to play the role of the typicality-defining measure in particularism. Later, however, in Section 7, we will show how they can be used to underpin an entirely different and novel approach to explaining the AoT.

\subsection{Dynamical similarity}\label{ssec:dynamical similarity}

We first give a definition of dynamical similarity. Suppose that a system's dynamics are determined by Hamilton’s principle, i.e. that its  DPMs $\gamma$ are those for which its action, $S$, is stationary:
\begin{eqnarray}
\delta S[\gamma]\Big|_\gamma = 0\,. \label{stationary}
\end{eqnarray}
A dynamical similarity in such a system is a transformation on the state space that rescales the action functional:
\begin{eqnarray}
S\mapsto cS\,. \label{rescale}
\end{eqnarray}
For any system of this kind, a dynamical similarity will map a DPM to another DPM, and is therefore a symmetry. This follows straightforwardly from the fact that the stationarity condition (\ref{stationary}) is invariant under (\ref{rescale}).

Symmetries of this kind can be constructed for a very general class of systems.\footnote{ See, for example, the multitude of finite dimensional models treated in \cite{sloan2021scale} and the extension to field theory developed in \cite{sloan2024dynamicalsimilarityfieldtheories}.} They occur when there is a general scaling of $q$, $p$, and $t$ such that the Hamiltonian, $H$, rescales such that the canonical action,
\begin{equation}\label{eq:S can}
    S[\gamma] = \int_t \lf( p \de q  - H \de t \rt)\,,
\end{equation}
rescales by $c$.\footnote{ Sometimes it might be necessary to rescale other dimensionful coupling constants in the theory according to the prescription developed in \cite{sloan2021scale}. } In the $N$-body problem, for instance, the scalings
\begin{align}
    \vec q_I &\to c^2 \vec q_I & \vec p_I &\to c^{-1} \vec p_I & t \to c^3 t\,  \label{eq:kepler ds}
\end{align}
induce a dynamical similarity, where $(\vec q_I, \vec p_I)$ are the position and momentum of the $I^\text{th}$ particle. The transformation \eqref{eq:kepler ds} ensures that $S \to c S$ for a $-1/|\vec q_I - \vec q_J|$ potential. For the Kepler problem (i.e, $N$-body problem with $N = 2$), \eqref{eq:kepler ds} is equivalent the well-known symmetry resulting from the conservation of the norm of the so-called \emph{Runge--Lenz} vector \citep{prince1981lie}.

Another dynamical similarity, more directly relevant to the goals of this paper, occurs in FLRW cosmology (i.e. homogeneous and isotropic cosmology) with a set of real scalar fields $\phi_i$, spatial curvature $k$ equal to zero and positive cosmological constant $\Lambda >0$. The phase space variables can be defined as: the spatial volume $v = a^3$ of a reference patch of space co-moving with respect to free-falling observers, its conjugate momentum $h = -3H$ (where $H$ is the Hubble parameter), the scalar fields $\phi_i$, and their conjugate momenta $\pi_i$. In these variables, the transformation
\begin{align}\label{eq:flrw ds}
    v &\to c v & \pi_i &\to c \pi_i
\end{align}
rescales the canonical FLRW action
\begin{equation}\label{eq:frlw action}
    S_\text{FLRW} = \int_t \lf[ \dot v h + \dot \phi_i \pi_i - N v \lf(  -\frac {h^2}2 + \frac{\pi^2}{2v^2} + \frac {3\Lambda}2 + U(\phi) \rt)  \rt]
\end{equation}
by $c$ and is therefore a dynamical similarity.

\subsection{Dynamical similarity and physical identity}\label{ssec:ds as gauge}

When should DPMs related by dynamical similarity be seen as physically identical? This question is highly relevant to the particularism/generalism debate concerning the AoT: The particularist project assumes that there is a preferred measure on the space of DPMs, which can be used to characterize low-entropy early states as highly atypical.

In the Kepler problem, where the two bodies are seen as a star and a planet orbiting it, the DPMs linked by dynamical similarity \eqref{eq:kepler ds} are two planetary orbits with different size and energy (but the same eccentricity and orientation). These are clearly physically different: The sizes and periods of these orbits can be compared to standard rulers (e.g., the radius of the Earth) and clocks (e.g., the sidereal day). These relative quantities, which are readily available to the experimenters that typically use Kepler's laws, give empirical means to distinguish transformed representations of the intended target systems.

On the other hand, in situations where the dynamical similarities extend to act on \emph{all} objects that the experimenters have access to (including all available clocks and rods), there is no way to detect the effects of the appropriately extended set of dynamical similarities. This is the situation we typically have in mind when modelling the matter distribution of the Universe. In this case, our knowledge about the relationship between the density fluctuations of the CMB and the large-scale structure of the Universe comes from large $N$-body simulations, like the Millennium Simulation \citep{springel2005simulations}, on an expanding FLRW background. If we grant the inability to observe the cosmic volume $v$ in an FLRW system (see below), then we should grant the inability to observe the global size of the $N$-body system expanding on it. In the limit where the expansion effects are small compared to the timescales involved in the $N$-body system, we can, therefore, treat \eqref{eq:kepler ds} as a gauge symmetry of the cosmological $N$-body system.

In the case of FLRW theory, it arguably \emph{is} natural to interpret the dynamical similarity \eqref{eq:flrw ds} as relating two physically identical DPMs. One powerful motivation for this perspective comes from the observation, first suggested in \cite{Ashtekar:2011rm} with subsequent developments in \cite{Sloan:2014jra,Sloan:2015bha,Sloan:2016nnx}, that in the equations of motion for this system, the spatial volume $v$ does not appear at all! These equations of motion, obtained by varying $v$ and $\phi$, are the Friedmann and Klein--Gordon equations:
\begin{align}
    \frac {h^2}2 &= \frac{\dot \phi^2}2 + \frac {3\Lambda}2 + U(\phi)
    & \ddot \phi_i + h \dot \phi_i + \diby{U}{\phi_i} &= 0 \,.
\end{align}
Since all cosmological observations depend only on the solutions to these equations and the variables in them, models connected by the dynamical similarity \eqref{eq:flrw ds} have no observable differences.\footnote{ The construction is a bit more subtle when $k\neq 0$ and $\Lambda \neq 0$ since there are too many dimensionful couplings in the original Hamiltonian to define the relevant dynamical similarity. One of these couplings must then be given a weight under dynamical similarity. See \cite{sloan2021new,sloan2023herglotz} for details of how to implement the construction under such a situation. Note, however, that current cosmological observations suggest that $k$ must be very small if it is different from zero \citep{baumann_2022}. } One may put this differently by saying that dynamical similarity is plausibly a \emph{gauge symmetry} in this context, since this is precisely how a gauge symmetry is usually understood: a transformation that leaves invariant the empirical content of the theory. A recently proposed systematic criterion for the identification of a gauge symmetry called the \emph{Principle of Essential and Sufficient Autonomy} (PESA) leads to the same conclusion \citep{Gryb:2021qix,gryb:phil_thesis}. Our proposal results from taking this conclusion seriously.

One corollary of this discussion of dynamical similarity is that whether one should count it as a gauge symmetry depends on the modelling context. The transformations \eqref{eq:kepler ds} should not be treated as gauge when modelling the position of planets in the solar system, but they \emph{should} be treated as gauge when modelling the position of galaxies in the Universe when the expansion effects are negligible. This context-dependence is also a general property of the PESA and results from the basic fact that models can only be expected to faithfully represent a target system when various modelling assumptions --- such as the idealisations, approximations and other auxiliary assumptions that depend on the context  --- are satisfied. Under such a view, the context-dependence is a general consequence of representing target systems by models and not an idiosyncratic feature of gauge symmetries.

\subsection{Treating dynamical similarity as gauge}

Suppose that we decide to treat some specific dynamical similarity as a gauge symmetry. How can we represent its dynamics purely in terms of physical quantities that actually do have physical meaning? For instance, how can we describe FLRW cosmology without referring to the (arguably unphysical) spatial volume $v$ in the first place? 

Gauge degrees of freedom are typically treated using a version of the so-called \emph{Gauge Principle}.\footnote{ See, for example, \cite{teller2000gauge} for a simple exposition of the Gauge Principle as applied to field theories. } Existing formulations of the Gauge Principle, however, do not apply to dynamical similarity. The technical reason is that a dynamical similarity must, by its definition \eqref{rescale}, rescale the symplectic potential $\theta = p\, \de q \to c p\, \de q$ in light of the definition of the canonical action \eqref{eq:S can}. But all existing formulations of the Gauge Principle assume that the dynamics of the theory can be written in terms of a fixed sympletic 2-from $\omega = \de \theta$ on the gauge-invariant subspace of phase space.

The details of how one can define a Gauge Principle for a general dynamical similarity are given in \cite{gryb:phil_thesis} and a similar construction has been provided by \cite{bravetti2022scaling}. The basic idea is to project the dynamics of the original symplectic system onto an invariant subsystem either by explicitly taking the quotient of the action with respect to the dynamical similarity (as is done in \cite{bravetti2022scaling}) or by projecting the dynamics onto a gauge-fixed surface transverse to the generator of dynamical similarity (as is done in \cite{gryb:phil_thesis}). The formal details are not important here. What is important is that the invariant subspace is \emph{not} a symplectic manifold, and therefore has no natural sympletic 2-form $\omega$. This can be seen in the FLRW theory discussed above by noting that the invariant variables $(H, \phi_i, \dot \phi_i)$ are an odd set (with $v$ removed) and, therefore, cannot form a symplectic manifold. Instead, the invariant dynamics is a flow on a so-called \emph{contact} space defined by a so-called \emph{contact 1-form} $\eta$. In the case of the FLRW theory, a simple choice for the contact form is $\eta_\text{FLRW} = \de h - v_i^\phi \de \phi^i$, where $v^\phi_i$ are the velocities of the scalar fields treated as independent variables.\footnote{ Note that there is an ambiguity in defining the contact form resulting from a symmetry of contact systems that are reparametrization invariant. See \cite{gryb:phil_thesis} for a demonstration of this symmetry and a discussion of how to understand the ambiguity. }

Contact spaces have a couple of features that will be important for understanding how our approach recovers key features of the AoT. First, the contact equations contain a term that can naturally be identified with a kind of friction-like force. More specifically, for a contact system with Hamiltonian $H$, coordinates $(A, q^i, p_i)$ and contact form $\eta = \de A - p_i \de q^i$, the contact equations are:
\begin{align}
    \dot q^i &= \diby{H}{p_i} \\
    \dot p_i &= - \diby{H}{q_i} - \diby{H}A p_i \\
    \dot A   &= - b H + \diby{H}{p_i} p_i\,,
\end{align}
where $b$ is a constant that, in our case, is chosen to be equal to the scaling dimension of the original Hamiltonian of the symplectic system.\footnote{I.e., $b$ is such that $\delta H = bH$ under an infinitesimal dynamical similarity.} The first two equations are the usual Hamiltonian equations with an additional contribution $- \diby{H}A p_i$ in the equation for $\dot p_i$. This contribution is normally interpreted as a linear drag term analogous to the drag term in, for example, a damped harmonic oscillator. We will call the prefactor $\diby H A$ of this term the \emph{drag}. It differs from the usual drag coefficient of damped systems, however, in that it can be a non-trivial function on contact space. The last line describes the dynamics of $A$, which, if non-trivial, can also affect the dynamics of the drag.

Note that the dynamical variable $A$ depends on the drag such that it renders the \emph{overall} set of equations time-reversal invariant. This is different from a standard damped system, where the damping term breaks time-reversal invariance because the usual drag coefficient is a constant and not a dynamical variable. Nevertheless, if the Hamiltonian of the contact system is such that the drag is monotonic, then the system can contain attractors due to the dissipation of energy within the system. We will define attractors more carefully in \Sec\ref{sub:attractors} below. These considerations indicate that, remarkably, contact systems can contain attractors even if the dynamics is time-reversal invariant. The consequences of this will be central to our general scheme for explaining the AoT.

The second feature of contact systems that will be relevant to our considerations is that while there is a natural time-\emph{independent} volume-form (i.e., $\omega^N$) on a $2N$ dimensional phase space, there is no natural time-\emph{independent} measure on a contact space. Instead, the natural volume forms $\text{vol}$ on a contact space satisfy
\begin{equation}
    \deby{\text{vol}}{t} = -n\diby H A \text{vol}\,,
\end{equation}
where $n$ is the dimension of the configuration space of the original symplectic system. That result means the volume-form $\text{vol}$ will contract (or expand) if the drag is non-zero. Thus, depending on the time-dependence of the drag, the resulting time-dependent measure on the space of states will upend the typicality arguments that motivate a PH. In fact, we will see that this is precisely what happens in $N$-body and cosmological models.

\section{The Janus-Attractor scenario as a third way}\label{sec:general JA}

\subsection{Intermezzo: a clue from the Boltzmann--Sch\"utz hypothesis}

To identify a way out of this dilemma, it is helpful to point out a common feature, and possible limitation, of both the generalist and particularist accounts of the AoT as envisaged by Price: they both seek to give an explanation of the AoT in terms of global features of the theory and its DPMs. Particularism, for example, postulates a particular fact in addition to time-reversal invariant laws to constrain the DPMs. Similarly, generalism postulates non-time-reversal invariant laws to constrain the DPMs. We will call such explanations \emph{external} because they depend on modelling constraints that are external to the target systems being studied.

It has long been pointed out, however, that there could also be non-external explanations of the arrow of time. Explanations involving \emph{internal} features of the target system --- such as contingent facts about particular observers --- are also possible. Here, we will refer to these as \emph{internal} explanations. One such example,  pointed out by \citet{cirkovic2002} not long after the publication of Price's paper \citep{price2002boltzmann}, is the so-called \textit{Boltzmann--Sch\"utz hypothesis}. \'{C}irkovi\'{c} dubs this approach \textit{acausal-anthropic} because we (say, human observers on Earth) are the particular observers in question.

The Boltzmann--Sch\" utz hypothesis is in some respects --- but importantly not all --- particularism without the requirement that the particular fact be an \emph{initial} condition. The hypothesis stipulates that the Universe (here with an upper case `U') is either infinite in spatial extent, or at least much larger than the visible universe. In modern terms, this implies an infinite collection of causally disconnected regions of spacetime that \cite{cirkovic2002} refers to as a \emph{multiverse}, although strict causal disconnection is not essential to the argument. All that matters is that there are vast regions of spacetime that observers like us have no direct knowledge about.

In the vast majority of such regions, entropy is locally high, near its maximal value. However, as is to be expected statistically due to the Universe being so incredibly large, there are occasional randomly distributed low-entropy fluctuations. The existence of complex lifeforms, by their very nature, is tied to these fluctuations or, more precisely, the entropy gradient that the fluctuations give rise to locally. The fact that we find ourselves in such a region is explained by the fact that, in virtue of our being forms of life, we could not possibly have found ourselves anywhere else. The relative atypicality of the AoT is then explained by the relative atypicality of regions that our existence is tied to.

This explanation of the AoT is \emph{anthropic} in the sense originating from \cite{dicke,Carter1974}. It can be seen as \emph{internal} as it relies on the existence of special observers internal to the Universe. According to it, the AoT is not a global feature: The vast majority of regions have no AoT, and there are isolated others that harbour agents for whom the local temporal gradient is reversed relative to ours. Indeed, we should expect the number of regions with our AoT to be roughly equal to the number of regions with the opposite AoT. To sum up, in the Boltzmann--Sch\"utz hypothesis, the AoT is interpreted as a mere internal affair: an unavoidable feature of our local environment on which our very existence is contingent.

Arguably, the Boltzmann-Sch\"utz hypothesis is unconvincing. Besides the fact that the objections to particularism discussed in \Sec\ref{sec:objs particularism} which rely heavily on the notion of entropy apply to it as well, the hypothesis also faces serious problems unique to it. First, according to well-known arguments, one might expect that ``most'' agents in low-entropy fluctuations do not actually have the low-entropy past encoded in their apparent memories. In other words, these memories are no reliable witnesses of their actual past, giving rise to a version of the so-called \emph{Boltzmann Brain} problem \citep{Price2013,Carroll2017-CARWBB-2}. Second, the entropy in the actual observable universe appears to be much lower than what is strictly needed for our existence --- counter to the spirit of the anthropic argument. Finally, to the extent that \'{C}irkovi\'{c} ties the Boltzmann--Sch\"utz hypothesis to a multiverse-like cosmic hypothesis, it faces the general methodological worries confronting multiverse theories, namely, that the prospects for empirically testing such theories seem bleak overall \citep{friederich2021}.

Even if we do not wish to adopt the Boltzmann--Sch\"utz hypothesis to the letter, we may still be motivated by some of its key ingredients. In particular, our own proposal will implement the following ideas:
\begin{enumerate}[A]
\item The AoT is a local, internal affair and not a global, external condition.
\item The time directions that we associate with past and future are determined by local (in cosmological time) entropy gradients that do not generalize universally. Indeed, the number of local arrows pointing in one time direction should be roughly balanced by the number of local arrows pointing in the opposite direction.\label{JA features: symmetric arrows}
\item What needs explaining is why an AoT is observed by agents like us in our current cosmological epoch.
\end{enumerate}
The Janus-Attractor scenario retains all of these ingredients. However, it differs from the Boltzmann--Sch\"utz hypothesis in that it reveals these observers \emph{not} to be tied to low-entropy fluctuations but to be ``close'', in a sense to be defined below, to a particular dynamical attractor. We will now describe the Janus-Attractor scenario by first describing its two central notions: attractors and Janus points. The structures defined by these notions are possible when we recognise dynamical similarity as a gauge symmetry and consider the dynamics on a contact, rather than a symplectic, state space.

\subsection{Attractors}\label{sub:attractors}

An attractor $A$ is a set of points on a manifold $M$ that is invariant under the dynamical flow induced by a function $f$ that defines time evolution. To eliminate uninteresting cases, an attractor is usually required to have a so-called ``basin of attraction,'' $B(A)$, which differs from $A$ by a set of non-zero measure. This basin of attraction is defined to be the set of points that flow into $A$ asymptotically.

A sharp definition of ``attractor'' is given in \cite{milnor1985concept}. Milnor's definition requires a measure $\mu$, in order to determine the existence of basins of attraction using the sets of nonzero measure, and the attractor is hence called a \emph{measure attractor}. The definition also requires a Riemannian metric on $M$ to define $A$ in terms of the asymptotic flow. However, importantly for what follows, the existence and identity of measure attractors are independent of the particular choice of measure or metric used.

Given a choice of Riemannian metric on $M$, we can define a distance function $d(x,y)$ between any two points in $M$. Then, for any $x \in B(A)$ and in the limit $t\mapsto +\infty$, the distance between $f(t ,x)$ and some set $\omega(x) \subset A$ will go to zero.\footnote{ The set $\omega(x)$ is called the $\omega$-limit set of $x$. } This allows us to talk about ``closeness'' to an attractor: a point $y$ is getting close to an attractor as its distance to the attractor goes to zero.

Some features of these definitions are worth taking note of. First, while the actual distance along an orbit to an attractor depends on the metric chosen, its vanishing in the limit is metric-independent. Second, the direction of \emph{increasing} $t$ and \emph{decreasing} $d(f(t,x), \omega(x))$ along the flow gives us a way to identify the time orientation on an orbit where the states $f(t,x)$ are ``approaching'' an attractor. Finally, while it is more common to study attractors in theories that are \emph{not} time-reversal invariant, contact systems allow for the possibility of having attractors in theories that are. In such cases, barring any global obstructions, if an attractor exists in the \emph{forward} (i.e., $t\to +\infty$ limit) flow of $f$, then there must also be an attractor in the \emph{backward} (i.e., $t\to -\infty$ limit) flow as well. Because of this, any dynamical trajectory that flows to an attractor in increasing $t$ also flows to an attractor in decreasing $t$, in line with our expectations from the feature \ref{JA features: symmetric arrows}.

\subsection{Janus points}

The concept of an attractor allows us to distinguish between time orientations that point towards and away from a particular attractor when starting from a point $x$ along a dynamical flow $f$. This is a necessary ingredient to identify a local AoT for an observer at $x$ in our scenario. But while an attractor can give us a way to distinguish between time orientations along an orbit, it cannot by itself give a good reason to align such time orientations with an AoT because, as discussed at the end of \Sec\ref{sub:attractors}, an unoriented orbit may start and end on two different attractors, and the symmetry between them gives no way to privilege any particular time direction. In addition, there is no natural way to say that a point $x$ is closer to one attractor than it is to the other without introducing a preferred metric on $M$. To have a chance at giving a full account of an AoT including the distinction between past and future directions, one therefore needs to break the symmetry between the endpoints of a curve and show that there exists a significant monotonic gradient in some quantity towards one of the endpoints in question. To do this, we will introduce an extra structure along an orbit of the flow that we will call a Janus point.

The idea of a Janus point was first introduced in \cite{Barbour:2014bga} and played a central role in a later book by Julian Barbour \citep{barbour2020janus}. In these works, the Janus point is understood as a state on an orbit from which point oppositely directed AoTs --- hence the reference to the two-faced Roman god Janus of beginnings and transitions. Beyond this intuitive definition, however, it is difficult to find an explicit formal characterisation of a Janus point. An early attempt to do so was given in \cite{Gryb:2021qix}, where a classification of Janus points was given to characterise certain differences between the Janus points encountered in cosmological and $N$-body systems. We will not need to make such a classification here and will attempt to provide a general definition that both implements the basic idea of a Janus point and works for all the different systems we want to consider. We should note, however, that our presentation may differ in letter, though not in spirit, from other earlier constructions in the literature.

The intuition underlying our construction is to understand a Janus point as a point $j$ along the orbit $\gamma$ of a real dynamical system where the flow is, at that very point, \emph{preserving} some measure $\mu$. Loosely speaking, at the Janus point the time-derivative of the measure, $\dot{\mu}$, goes through zero in such a way that there exist two oppositely directed time orientations on $\gamma$ where $\dot{\mu} > 0$ on either side of $j$.\footnote{Technically, this can be expressed by characterizing a \textit{Janus surface} $\mathcal J$ as a codimension-1 submanifold of $M$ transverse to the generators of the flow of $f$ such that
\begin{eqnarray}
\frac{\de \mu[f(t,\mathcal J)]}{\de t}\Big|_{t=0}=0\,,
\nonumber
\end{eqnarray}
where $\frac{\de \mu[f(t,\mathcal J)]}{\de t}$ is the more formal expression of what is referred to as ``$\dot{\mu}$'' in the main text. A Janus point is any point $j \in \mathcal J$ in a Janus surface.}

Note that, according to how we have characterized Janus points, their identity depends on a particular choice of measure $\mu$. Different choices of $\mu$ will generally yield different collections of Janus points. This introduces a certain degree of conventionality into the notion of a Janus point. This conventionality does not exist when defining attractors alone, since our definition of attractor did not depend on any particular choice of measure.

\subsection{The Janus-Attractor scenario}\label{sec:JA scenario}

In this section, we give a general scenario for establishing an AoT using the concepts of an attractor and a Janus point. The idea will be to consider an observer that attributes a state $x$ to the current state of the world. We then require the following conditions, which define a Janus-Attractor (JA) scenario:
\begin{enumerate}[i]
    \item $x$ is in the forward flow of some Janus point $j$.
    \item $j$ (and therefore $x$) are in the attractive basin $B(A)$ of an attractor $A$.
    \item The distance $d$ between $x$ and $\omega(x) \in A$ tends towards zero as $t \to +\infty$.
\end{enumerate}
These conditions entail that $x$ is on an orbit of the flow with a time orientation such that a Janus point is in the backward flow of $f$ and the attractor is in its forward flow. Moreover, $x$ must be close, in the sense defined by the distance function $d$, to an approaching attractor in the direction of increasing $t$. When these conditions are satisfied, we contend, there is an AoT pointing in the direction of increasing $t$ from $j$ to the set $\omega(x) \in A$.

\begin{figure}
    \centering
    \includegraphics[width=0.7\textwidth]{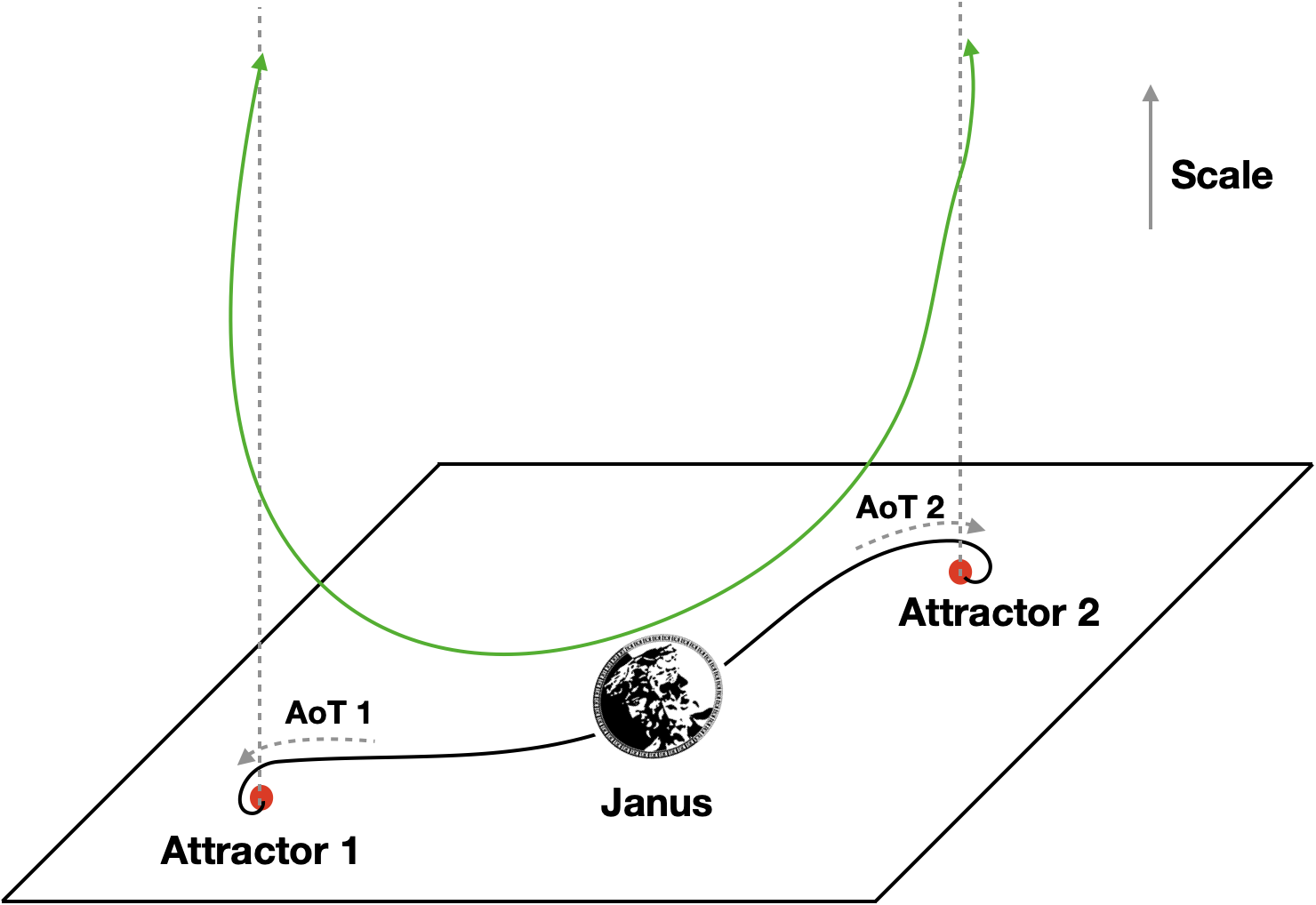}
    \caption{The Janus-Attractor scenario obtained by removing scale. Observers in states near an attractor experience an Arrow of Time pointing from the Janus point to the nearby attractor. In a time-reversal invariant theory, the arrow experienced by an observer in the forward flow is balanced by an oppositely oriented arrow experienced by an observer near the attractor in the backward flow. }
    \label{fig:JA scenario}
\end{figure}

In such a situation, the Janus point defines a time orientation for certain classes of observers because it locally breaks the time-reversal invariance of $\gamma$ that exists when $\gamma$ is bounded by two attractors. Given such a time orientation, the requirement that $x$ be close to $A$ implies that there is a quantity, the inverse of the distance to the attractor, $d^{-1}(x, \omega(x))$, that is large and increasing monotonically. This quantity formally defines a general AoT in the sense specified in the Introduction (\Sec\ref{sec:intro}). Note that, for every AoT experienced by an observer near $A$, time-reversal invariance implies the existence of a second AoT with the opposite orientation experience by an observer near the attractor in the backward flow. The overall picture is illustrated in Figure~\ref{fig:JA scenario}.

What we need to argue for now is why the inverse of a distance function $d$ might define a physically interesting AoT. One reason to believe that it could be is that, as we have defined them, attractors are fixed sets of dynamics. They therefore correspond to a kind of equilibrium set of the system. The stability of such equilibrium sets is guaranteed when one can find a so-called \emph{Lyapunov variable} \citep{lyapunov1992general} in the system. A Lyapunov variable is a function whose gradient along the flow is always negative except at $\omega(x)$, where it tends to zero. A function bounded by the distance $d(x,\omega(x))$ is a special class of Lyapunov variable (see Definition~5 of \cite{beretta1986theorem}). Whenever a Lyapunov variable exists, one can always construct a function $S(x) =- L(x)$ that increases monotonically along the flow and reaches its maximum near equilibrium. These functions have been identified with non-equilibrium entropy functions in \cite{beretta1986theorem}. When there is a JA scenario, a natural candidate for such an entropy function $S(t)$ is simply the distance from the Janus point to the current point $x$; i.e., $S=d(j,x)$. In this case, $S=0$ at the Janus point and grows monotonically until it reaches its maximum value at equilibrium.

Closeness to an attractor is thus a reasonable notion for defining an AoT since such an AoT is aligned with the temporal direction in which the system is equilibrating such that a natural entropy function can be defined. These considerations give us general reasons to believe that an AoT with certain acceptable physical characteristics will arise in a JA scenario. However, there is no guarantee that for any particular physical system there will be choices of measure, metric, and entropy function that are simple and physically interesting.

One hopeful possibility, which is explicitly realised in the models we consider in the next \Sec\ref{sec:JA and AoT}, is that the drag of the contact system obtained by applying the Gauge Principle to dynamical similarity will behave like $S = d(j,x)$. When this happens, the drag will increase monotonically from the Janus surface and will reach its maximum value at the attractor. Measures with such properties are potentially interesting for a variety of reasons. First, they highlight the geometric features of the state space in such a way that (i) a Janus point appears as a point where there is no focusing of solutions and (ii) attractors are maximally focused sets. This could be useful if the Janus points and attractors have interesting physical significance within the theory. Conversely, if the focusing itself can be understood in more physical terms, then that understanding could be used to explain the physical significance of the Janus points and attractors.

These considerations show that a JA scenario locally exhibits time-directed behaviour reminiscent of an AoT. However, we have not yet shown that the rich AoT phenomenology in our Universe can indeed be explained through a JA scenario. Having moved from phase space to contact space we can no longer encode that phenomenology in terms of a Past Hypothesis \citep{ph_takedown}. In what follows, we propose two general features of our Universe that entail and encode the AoT phenomenology -- or at least parts of it -- without relying on a Past Hypothesis. These are, namely, the fact (i) that the early Universe seems to have been extremely smooth and the fact (ii) that the more distant an object is, the more we see it red-shifted. We will later reproduce these features using simple models that exhibit JA scenarios.

\section{Narrowing down the explanatory target}\label{sec:explanatory target}

In the absence of a time-independent measure on DPMs, no preferred notion of entropy exists, and hence we cannot characterize the problem of accounting for the AoT in terms of accounting for the phenomenology associated with entropy increase. However, as we argue in this section, the challenge of accounting for the AoT can be reframed in terms of two more specific empirical problems that beset modern cosmology. Later we will argue, in \Sec\ref{sec:JA and AoT}, that JA scenarios offer excellent prospects for resolving those problems, thereby effectively solving the AoT problem as well.

The two problems are: 
\begin{enumerate}[(1)]
	\item \textit{The smoothness problem:} Why was the Universe so remarkably smooth early in its history?\label{prob:smoothness}
	\item \textit{The red-shift problem:} Why was the Hubble parameter, which measures the amount of red-shifting in the Universe, so large and monotonic in the past? \label{prob:red_shift}
\end{enumerate}
In what follows, we elaborate upon these questions and explain why good answers to them go a long way towards providing a convincing account of the AoT.

\subsection{The smoothness problem}

By saying that the early Universe was extremely smooth we mean that the distribution of energy-momentum and geometric degrees of freedom was approximately homogeneous and isotropic in it. The call to explain that smoothness is familiar from the particularist literature. \cite{price2004origins} argues:
\begin{quote}
The crucial thing is that matter in the Universe is distributed extremely smoothly, about one hundred thousand years after the Big Bang. 

...In effect, the smooth distribution of matter in the early provides a vast reservoir of low entropy, on which everything else depends. The most important mechanism is the formation of stars and galaxies. Smoothness is necessary for galaxy and star formation, and most irreversible phenomena with which we are familiar owe their existence to the sun. [p. 227-228, original italics]
\end{quote}
The motivation here is that smoothness leads to a very low-entropy state when the system is self-gravitating. Because the force of gravity is attractive and leads to the clumping of point masses, smooth states that are not clumpy are entropically suppressed.\footnote{ More specifically, clumpy states occupy large phase space volumes because the $-1/r$ gravitational potential is peaked on them. } The observation that the contents of the universe were very smooth at a particular point in its evolution is then taken to imply that the universe was in a low-entropy state at that time.

These arguments have a long history.\footnote{ See \cite{PADMANABHAN:1990book} for detailed derivations of gravity's unusual thermodynamic properties and how the statistical mechanics of the gravitational $N$-body problem can be applied to galaxies. } Their relevance to the AoT was prominently brought to the attention of the philosophical community by \cite{Penrose:1979WCH}, see also Chapter 7 of \cite{Penrose:NewMin}. In reference to those arguments, \cite{price2004origins} goes as far as declaring that the ``discovery about the cosmological origins of low entropy[,] is the most important achievement of late twentieth century physics.'' According to him, this observation is the \emph{only} thing that needs explaining in order to account for the thermodynamic AoT because it entails ``most irreversible phenomena with which we are familiar.''

Let us evaluate this claim.

To begin with, it is true that, in order to even apply thermodynamic notions to the universe, we must already take advantage of the observational fact that it is, to a good approximation, relatively smooth on large scales --- at least during the epochs to which we have direct observational access. To understand why, consider a patch of space containing smooth matter whose boundary evolves as if it were made of free dust particles. Such a patch of space acts as a kind of box through which there is effectively no heat flow because the homogeneity of the matter distribution implies that any heat flowing out of the box should be balanced by an equal amount flowing in. This allows us to define the thermodynamic properties of the matter in that box, provided a variety of physical conditions on the states of the matter are met.\footnote{ For example, the timescales of changes in the temperature and other thermodynamic quantities must be large compared to the mixing times of the matter. } If we take the box to be the spatial boundary of the visible universe; i.e., the finite part of the whole `Universe' that is causally accessible to us (which we will henceforth call the `universe' with no capitalization); talk about the temperature and entropy of matter (or radiation or geometry) becomes possible. Thus, smoothness is a necessary, but not sufficient, condition for defining the entropy of the universe in the first place.

There is strong evidence that the early universe was indeed very `smooth.' The epoch of $10^5$ years after the Big Bang, mentioned by Price, is relatively late in the evolution of the early universe. This is roughly the time of \emph{recombination} in which the universe cooled to the point where the first electrically neutral hydrogen atoms could form, releasing the first visible light in the universe called the \emph{Cosmic Microwave Background (CMB)} radiation.\footnote{ More precisely, recombination occurred $3.78\times 10^5$ years after the Big Bang. It's not completely clear whether Price meant this period or some slightly earlier epoch. } The fact that any cooling occurred at all is an essential assumption of Price's claim, and, as we argue below in \Sec\ref{ssec:red-shift problem}, does not follow from early smoothness alone. But, putting this assumption aside for now, the universe indeed seems to have been relatively smooth at the time of recombination. This evidence comes from direct observations in radioastronomy of the CMB, indicating temperature variations of $1$ part in $10^5$ over background levels.\footnote{ The details about this and other cosmology evidence discussed in this section can be found in any good cosmology text book such as \cite{dodelson2020modern}, \cite{mukhanov2005physical}, or \cite{weinberg2008cosmology}. Our own treatment roughly follows \cite{baumann_2022}. See Chapter~7 of \cite{baumann_2022} for an introduction to CMB observations.}

It is interesting to question how early the universe was smooth and how smooth it was before the time of recombination. However, these questions are not particularly relevant when it comes to explaining the AoT today, over and above the question of why the universe was smooth at the time of recombination. The inflationary paradigm of modern cosmology (which to a certain extent is still speculative) may shed light on this, but its explanatory power with respect to early smoothness remains controversial.

Once smoothness at the time of recombination is assumed, clumping due to the self-gravitation of (comparatively) slightly overdense regions amplified inhomogeneities that were small at recombination, resulting in the formation of dense gas clouds and, ultimately, stars. (See the Millennium Simulation \citep{springel2005simulations}). First generation stars collapsed under gravity to produce later generations of stars including our Sun --- a vast reservoir of entropy that we understand to be the source of nearly all the time-asymmetric processes on Earth. However, as we will see in the next section, this vast reservoir of entropy is \emph{not} completely, or even mostly, due to the gravitational collapse of the CMB inhomogeneities.

\subsection{The red-shift problem}\label{ssec:red-shift problem}

\subsubsection{Red-shift and the cosmological arrow}

Most of the entropy that ultimately powers time-asymmetric processes on Earth is generated in stellar nuclear fusion, not in the gravitational collapse that followed recombination. A back-of-the-envelope calculation shows that the fusion entropy dwarfs the collapse entropy \cite{wallace2010gravity}. Gravitational collapse therefore acts mainly as a catalyst: by compressing protostellar gas it initiates fusion, thereby unlocking the huge ``entropy reservoir'' stored in stellar nuclear fuel whose dissipation drives the local arrow of time.

The existence of that reservoir is traced to the extraordinarily high \emph{rate of expansion} in the early Universe. Observationally, expansion appears as the systematic red-shifting of light, which is proportional to the Hubble parameter $H$. What matters for entropy production is not the absolute size of the Universe but the magnitude of $H$, consistent with our conclusions from \Sec\ref{sec:DS as gauge} that cosmological observations depend only on the relative rate of change of the size of the Universe and not its absolute scale. Standard arguments give the temperature of the early universe $T\propto\sqrt{H}$ \cite[Eq. 3.55]{baumann_2022}, so as $H$ falls, the cosmic temperature drops.

Because the microscopic collision timescale $t_c$ lengthens while the expansion timescale $t_H\sim1/H$ shortens, epochs inevitably arise when $t_c\gtrsim t_H$. At those “freeze-out’’ moments the rapid expansion prevents further reactions, locking in out-of-equilibrium particle abundances. A sequence of such freeze-outs punctuates cosmic history and demarcates its major epochs, as we will revisit in below in \Sec\ref{sec:global to local arrow}. Each freeze-out is intrinsically time-asymmetric: had the Universe begun in global equilibrium, none could have occurred. Thus, the monotonic decline of $H$—and the huge initial red-shift it implies—provides the missing nonequilibrium ingredient that both explains the freeze-out moments and accounts for the vast entropy reservoirs later present in stars.

\subsubsection{From a cosmological to a local arrow}\label{sec:global to local arrow}

During Big Bang Nucleosynthesis (BBN), which began $\approx  3$ minutes after the Big Bang, the cooling universe first made deuterium nuclei stable. This triggered a burst of fusion that locked most available neutrons into helium. Had expansion halted, fusion would have continued until virtually all hydrogen was consumed, leaving no fuel for future stars. Instead, the Hubble rate $H$ kept falling, quenching nuclear reactions early and preserving a vast hydrogen surplus: an embryonic entropy reservoir whose later release in stellar fusion powers time-asymmetric phenomena on Earth \citep{wallace2010gravity}.

That surplus hinged on a prior ``neutron bottleneck.'' Before BBN, neutrons and protons were held in near-equal numbers by beta and inverse-beta processes
\begin{align}\label{eq:beta decay}
n+\nu\_e &;\leftrightarrow; p^{+}+e^{-}\\
n+e^{+} &;\leftrightarrow; p^{+}+\bar{\nu}\_e , .
\end{align}
Because neutrons are slightly heavier, cooling tipped the balance toward protons; when the weak interaction froze out, the neutron fraction ``froze in'' at about $1/8$. Subsequent neutron decay further reduced their number, capping helium production at roughly one atom per four hydrogens and leaving enough stellar fuel for galaxies like ours to form.

BBN illustrates a general pattern: each cosmological epoch ends when the ever-slowing expansion pushes the interaction timescale $t_c$ above the expansion timescale $t_H\sim1/H$. These freeze-out moments repeatedly kick the universe out of any path toward global equilibrium, trapping entropy in metastable reservoirs—initially neutrons, later hydrogen, and eventually the chemical free energy that planets exploit. Understanding how $H$ monotonically declines is therefore central to explaining the origin of the cosmic and terrestrial arrows of time.

Tracing the chain further back reveals earlier freeze-outs: the hadron epoch ($\approx 10^{-5}$ s) when quarks bound into nucleons; the quark–gluon plasma of the preceding quark epoch; electroweak symmetry breaking; and, more speculatively, inflation and reheating. At still higher energies, quantum-gravity effects dominate. Penrose estimates that a maximal-entropy black-hole state in this regime would be $10^{10^{120}}$ times likelier than the smooth universe we see \cite[Ch. 7]{Penrose:NewMin}. He therefore proposes the Weyl Curvature Hypothesis, positing vanishing Weyl curvature at the Big Bang to enforce smoothness \cite{Penrose:1979WCH}; but because the Weyl tensor is conformally invariant, this hypothesis constrains geometry, not the value or monotonic fall of $H$.

The explanatory thread thus always returns to the red-shift (expansion-rate) problem. A satisfactory account of the thermodynamic arrow of time must show why the early Universe began both extraordinarily smooth \emph{and} with a huge, steadily decreasing $H$. Only then can it explain the cascade of entropy reservoirs—from excess neutrons to surplus hydrogen to stellar fusion, which ultimately drives the irreversible processes we observe on Earth \citep{Rovelli:2018vvy}.

\section{Accounting for the AoT with a JA scenario} 
\label{sec:JA and AoT}

When we introduced the JA scenario, we noted that it contains a natural candidate for a surrogate entropy function $S(t)$, namely, the distance $d(j,x)$ from the Janus point $j$ to the point x=$f(t,j)$, i.e. $S=-d(j,x)$. However, this does not show that a JA scenario can reproduce the rich AoT phenomenology around us. In this section, we partly fill this gap. We do this by outlining how JA scenarios arise in two concrete models that separately provide solutions to the smoothness and red-shift problems. We treat the red-shift problem first and the smoothness problem second.

\subsection{Explaining the red-shift problem in the JA scenario}\label{ssub:conc_red_shift}

The red-shift problem, as stated in \Sec\ref{ssec:red-shift problem}, demands an explanation for the rapid cooling associated with the monotonic decrease of the Hubble parameter that resulted in the low-entropy reservoirs responsible for many aspects of the AoT today. Here, we show that the model introduced at the end of \Sec\ref{ssec:dynamical similarity} ---  an FLRW cosmology with a single scalar field, positive cosmological constant and flat spatial curvature (i.e., $k=0$) --- reproduces this observed behaviour of the Hubble parameter. We choose this model for simplicity. As we will see, our conclusions robustly generalise beyond it. For more details about this model and for the claims made in this section, see Chapter~8 of \cite{gryb:phil_thesis}.

To solve the red-shift problem using a JA scenario we must show:
\begin{enumerate}[(i)]
    \item that the Hubble parameter decreases monotonically,\label{monotonicity}
    \item that subsystems generally approach an attractor, and  \label{attractor}
    \item that there is a function which one can identify with the drag in a JA scenario that places a Janus surface on or near the time slice playing the role of a ``Big Bang.''\label{janus}
\end{enumerate}
We now show that each of these points hold for the model of \Sec\ref{ssec:dynamical similarity}.

The phase space degrees of freedom of the model are the volume $v$ of a spatial slice, its conjugate momentum $h = -3H$, the scalar field $\phi$ and its momentum $\pi$. From the action \eqref{eq:frlw action}, the Hamiltonian is
\begin{equation}
    H_\text{FLRW} = N v \lf(  -\frac {h^2}2 + \frac{\pi^2}{2v^2} + \frac {3\Lambda}2 + U(\phi) \rt)\,,
\end{equation}
where $U(\phi)$ is the scalar field potential. The time evolution of $h$ is then
\begin{equation}
    \dot h = \pb h {H_\text{FLRW}} = - N \mathcal H + \frac{N \pi^2}{2 v^2} = \frac{N \pi^2}{2 v^2} \geq 0\,.
\end{equation}
Here we use the fact that the Hamiltonian constraint $\mathcal H = -\frac {h^2}2 + \frac{\pi^2}{2v^2} + \frac {3\Lambda}2 + U(\phi) = 0$ vanishes when the classical equations of motion are satisfied. Because the lapse $N$ is greater than zero for increasing $t$ and because $h$ is proportional to the \emph{negative} of the Hubble parameter $H$, we have
\begin{equation}
    \dot H \leq 0\,
\end{equation} 
for increasing $t$. This proves \ref{monotonicity}, namely, that the Hubble parameter is monotonically decreasing. 

To prove \ref{attractor}, we assume that the Weak Energy Condition holds for the scalar field. This translates into:
\begin{equation}
    \frac{\pi^2}{2v^2} + U(\phi) \geq 0\,.
\end{equation}

The Hamiltonian constraint $\mathcal H = 0$ then entails $H^2 \geq \frac {\Lambda}3$ (recall that $h = -3H$). The Hubble parameter $H$ starts from a large value,\footnote{In fact, this value can be shown to be $+\infty$ using a standard line of reasoning following the Hawking--Penrose singularity theorems \cite{penrose1965gravitational,hawking2023large}. } monotonically decreases, and then asymptotically falls to the limiting value $H = \sqrt{\frac \Lambda 3}$, which is its value during de~Sitter expansion. Because the solution approaches de~Sitter spacetime when $t\to+\infty$, de~Sitter spacetime is an attractor of the theory in the same temporal direction in which $H$ is monotonically decreasing, proving \ref{attractor}.

The result that de~Sitter spacetime is a late-time attractor of homogeneous cosmology was first proved in \cite{wald1983asymptotic} for a broader class of models than considered here. More generally, the \emph{cosmic no-hair conjecture} stipulates that, under general circumstances (usually involving one or more of the standard energy conditions), general relativistic spacetimes with a positive cosmological constant have a de~Sitter attractor.\footnote{For a modern statement under general conditions, see \cite{andreasson2016proof}. For an introduction aimed at philosophers of physics, see \cite[Chap VII, \S 7]{belot2023accelerating}. For a more detailed analysis of the philosophical points, see \cite{doboszewski2019interpreting}.} Thus, it seems plausible that \ref{attractor} generalizes robustly beyond the specific idealised model considered here.

Finally, to establish \ref{janus}, notice that the velocity of the scalar field, $v_\phi \equiv \frac{\pi}{v^2}$, obeys Hamilton's second equation 
\begin{equation}
    \dot v_\phi = - N \lf( \diby{U}{\phi} - h v_\phi \rt)\,,
\end{equation}
where $Nh v_\phi$ plays the role of a drag term. If we choose a time parameter for which the lapse is chosen to be $N = 1/h^2$,\footnote{Recall that the proper-time $\de \tau = N \de t$ so that a choice of lapse fixes the time parameter or, alternatively, the choice clock for the evolution.} then the drag term reduces to $\tfrac 1{3H}$, which is zero at the Big Bang, where $H\to +\infty$, and maximal at the attractor. Thus, for such a choice of clock, the inverse of the Hubble parameter is a natural (monotonic) entropy function, $S(t)$, with the big bang as a Janus point.\footnote{ More concretely, the entropy function $S(t) = \frac 1 {3H} - \frac {1}{\sqrt{3 \Lambda}}$ is the negative of a distance function from the attractor taking its minimum on the Janus point and zero on the attractor. } This fulfils the requirements of the JA scenario for there to be an AoT pointing from the Janus point to the de~Sitter attractor. Since we know from observational cosmology that the universe is currently approaching de~Sitter-like expansion, this AoT in the model matches the actual AoT as far as the redshift problem is concerned.

To end this section, let us explore the characteristics of the Janus point in this picture. It is a straightforward calculation to show that all velocities in Hamilton's equations are bounded using the lapse $N = 1/h^2$, provided $v$ is excluded from the phase space as required by the considerations of \Sec\ref{ssec:ds as gauge}.\footnote{ One also has to assume that $\frac 1 {H^2} \diby{U}\phi < \infty$ when $H \to +\infty$ for this to hold. See \Sec 8.4.3 of \cite{gryb:phil_thesis} for more details. } Thus, for this clock choice, the Big Bang is a drag-free, Lipschitz continuous point of the dynamics and, as such, a Janus point. In \cite{KOSLOWSKI2018339,Sloan:2019wrz}, a similar result was shown to hold for more general models including Bianchi~IX models and more general FLRW theories.

We have thus established that the FLRW model exhibits a JA scenario with all the ingredients necessary for solving the red-shift problem, and we have shown that this finding generalises beyond the simplified constraints of that model.

\subsection{Explaining early smoothness in a JA scenario}\label{sec:explaning smoothness}

To solve the smoothness problem, we show that the $N$-body Newtonian system of self-gravitating point particle with positive energy introduced in \Sec\ref{ssec:dynamical similarity} exhibits a JA scenario and provide an argument showing that, in this model, states close to the Janus point are, relatively speaking, smoother than those close to an attractor. Finally, we sketch a further argument, with details to be found in \cite{gryb:phil_thesis}, which establishes that states near the Janus point are also smooth in an absolute sense.

Models like the $N$-body Newtonian system of self-gravitating point particle with positive energy are meant to approximate the relevant features of state-of-the-art simulations of structure formation in the early Universe, such as the Millennium Simulation by \cite{springel2005simulations}, which are essentially $N$-body models coupled to an expanding cosmological background. Because the expansion effects are minimal during the timescales of interest near the Janus point and homogeneous, we expect that the Newtonian limit is not in any way special as far as the ``smoothness'' of particle distributions is concerned. The positivity of the energy is meant to mimic the effect of a positive cosmological constant, which prohibits the matter distribution from recollapsing.

First, let us establish that the scale-free $N$-body system with positive energy has attractors for large-$N$ forward evolution. Theorem~1 of \cite{marchal1976final} states that, in such systems, the centre-of-mass coordinates $q^i$ (for $i = 1, \hdots, 3N$) behave as
\begin{equation}
    q^i \to A^i t + \mathcal O\lf(t^{2/3}\rt)
\end{equation}
as $t \to \infty$, where $A^i$ are (possibly zero) constants.\footnote{ This behaviour excludes the possibility of super-hyperbolic escape, which occurs in a set of measure zero on the space of solutions.} We can remove the scale by diving by the square root of the moment of inertia scalar
\begin{equation}
    I = \sum_{i} m_{ij} q^i q^j\,,
\end{equation}
where $m_{ij}$ is the mass matrix for the $N$-body system. Doing this, we find that the scale-invariant quantities $\hat q^i = q^i/\sqrt{I}$ obey
\begin{equation}
    \hat q^i \to B^i + \mathcal O\lf( t^{-1/3} \rt)\,,
\end{equation} 
for some new constants $B^i$ determinable from $A^i$. This means that, for large $t$, the scale-invariant $N$-body problem segments into subsystems that grow at most as $\mathcal O\lf( t^{-1/3} \rt)$, which shrink to the values $B^i$ at large times. The configurations $\hat q^i = B^i$ are, therefore, attractors of the $N$-body system. So while the scale\emph{full} configurations grow linearly in time, the scale-\emph{invariant} configurations approach attractors.

After establishing the presence of attractors, we can now sketch how a Janus surface arises. The $1/r$ dependence of the Newtonian potential the $N$-body model gives rise to the dynamical similarity
\begin{align}\label{eq:ds N body}
    t &\to c^3 t & q^i &\to c^2 q^i & p_i &\to c^{-1} p_i\,,
\end{align}
where $p_i$ are the momenta conjugate to $q^i$. We can eliminate this dynamical similarity by, as a gauge choice, fixing the value of $I$, which rescales under \eqref{eq:ds N body}, to some specific value. At the same time, we retain its time derivative $\dot I$ as an independent variable. We can then create quantities that are invariant under \eqref{eq:ds N body} by diving by the appropriate powers of $I$. In particular, the quantity
\begin{equation}
    S = \frac{\dot I}{I^{1/4}}
\end{equation}
is invariant under \eqref{eq:ds N body} and depends on the change of scale $\dot I$. $S$ is, thus, a natural invariant variable for encoding $\dot I$. A well-known result due to Lagrange and Jacobi\footnote{ See, for example, Equation~1.12 of \cite{marchal1976final}. } for $N$-body dynamics with positive energy says that $\ddot I = T + E \geq 0$, where $T$ is the total kinetic energy of the system and $E$ is the total energy. This, combined with the positivity of $I$ and the relative largeness of $T$, leads to the result that
\begin{equation}
    \dot S \geq 0\,.
\end{equation}
This suggests that $S$ can be treated as an entropy function. Indeed, in \cite{gryb:phil_thesis} it is shown that $S$ is proportional to the drag function arising when using the gauge fixing condition that keeps $I$ fixed. This means that the system has a natural Janus point when $S = 0$.

One can see that states near the Janus point must, in an intuitive sense, be ``smoother'' than states near attractors by considering energy conservation: (where we now restrict to zero energy for simplicity)
\begin{equation}
    \frac 1 2 \sum_i m^{ij} p_i p_j + V(q) = 0\,,
\end{equation}
where $V(q)$ is the Newtonian gravitational potential. Note that, by diving by powers of $I$, we can split the kinetic term into a part that depends on the scale-invariant momenta $\hat p_i$ and another part that depends solely on $S$, so that one obtains:\footnote{ More concretely, $\hat p_i = I^{1/4}\lf( p_i - \frac {q^j p_j}I m_{ik} q^k  \rt)$. See \cite{gryb:phil_thesis} for more details. }
\begin{equation}
    \frac 1 2 \sum_i m^{ij} \hat p_i \hat p_j + \frac 1 8 S^2 + V(\hat q) = 0\,.
\end{equation}
Because $S$ is monotonic, it increases indefinitely towards the attractors as $t \to \pm\infty$. Since $\frac 1 2 \sum_i m^{ij} \hat p_i \hat p_j \geq 0$, $-V(\hat q)$ also grows indefinitely. But $-V(\hat q)$ growing indefinitely, intuitively, requires ``clumping'': it can only get large either when particles come close together so that $1/r \to \infty$ or when $I$ gets large due to the ejection of particles or subsystems. Either way, the inter-particle separations within a cluster must get small relative to the average size of the system. And this, intuitively, indicates ``clumping''. In other words, as the system approaches an attractor (i.e., as $t\to\pm\infty$), clumping increases dramatically. Conversely, by comparison earlier states will appear smooth to late-time observers.

Whether Janus point states are homogeneous in a sharp and absolute sense --- and not just intuitively and in relation to later states --- can be considered only in relation to specific measures on state space. Such considerations go beyond the scope of this work, but we can sketch the general procedure to establish this. Having chosen a particular gauge-fixing to eliminate dynamical similarity \eqref{eq:ds N body} one projects the Liouville measure for the $N$-body system onto this gauge fixing. Then one can use this projected measure to produce distributions of states at the Janus point and compare such states to distributions of $N$-particles generated by a uniform distribution.

Two candidate gauge-fixings have been considered in the literature. The first, given in Appendix~A.1 of \cite{barbour2013gravitational}, fixes the norm of $p$ and results in a \emph{completely} uniform distribution at the Janus point. The second, given in \Sec 8.3 of \cite{gryb:phil_thesis}, fixes $I$ (the norm of $q$), and leads to a \emph{near-uniform} distribution.

Taken together, these results indicate strongly that the $N$-body gravitational system exhibits a JA scenario in that `early' states near the Janus point are nearly smooth while `late' states near an attractor are highly clumped. This indicates that observers near highly clumped states will typically see smooth, nearly uniform, states in their past, meaning that the smoothness problem is solved to the extent that this model captures the key relevant features of our own Universe.

\section{Reconsidering the objections against generalism and particularism}\label{sec:reconsidering objections}

The present paper has its limitations, some of which will be mentioned in the concluding section. However, being neither generalist nor particularist, it avoids all the objections against generalist and particularist approaches that we presented in \Sec\ref{sec:the_dilemma}. Here we address them one by one.

\subsection{Objections to generalism}

\begin{enumerate}[G.I]
    \item {\bfseries Objection from redundancy} This objection says that postulating time-asymmetric dynamics, as the generalist does, is redundant because a low-entropy past must be postulated anyway. The present approach deflects this objection because in the JA-scenario, on the one hand, locally time-asymmetric solutions are obtained automatically by eliminating dynamical similarity and, on the other, observers typically find that their past appears to be characteristically `low entropy' in the sense of being highly redshifted and comparatively homogeneous.
    \item {\bfseries Objection from lack of independent motivation}  According to this objection, the radical move from time-symmetric to time-asymmetric laws should be made only if it is also motivated by strong empirical reasons, not just by abstract qualms about the tension between time symmetry and the AoT. The objection thus challenges the move of changing the fundamental law. However, eliminating the cosmic scale factor from the formulation of cosmological theory and thereby removing dynamical similarity should be seen as clarifying, not changing, the fundamental law. Hence, this objection does not apply to the account of the AoT based on JA-scenarios.

    \item {\bfseries Objection from historical progress} This objection points out that, historically, our understanding has moved towards seeing time as an aspect of space-time and away from seeing it as inherently directed. Reassuringly, to JA-scenario-based account of the AoT can be seen as nicely fitting the overall historical arc wherein our understanding of time is no longer inherently directed: the interplay of attractors and Janus points creates a \emph{local} and emergent AoT, making ``past'' and ``future'' local and perspectival. One may compare this to how ``up'' and ``down'' have come to be understood as local and perspectival in the transition from the geocentric to the heliocentric picture of the world.
\end{enumerate}

\subsection{Objections to particularism}

\begin{enumerate}[P.I]
    \item {\bfseries Objections from mathematical and conceptual ambiguity} These objections focus on difficulties in identifying coherent and mathematically precise notions of entropy and typicality to states of the Universe. Such notions, however, are needed to argue that, once the assumption of a PH is made, entropy increase and, with it, the phenomenology of the AoT, is to be expected and not in need of any further explanation.

    In a JA scenario, the local AoT arises from the interplay of attractors and Janus points, and is not fundamentally characterised in terms of entropy. Attractors can be given precise mathematical definitions in terms of the $\omega$-limit sets of the dynamical flow as outlined in \Sec\ref{sub:attractors}. Similarly, Janus points can be defined in terms of the level surfaces of smooth functions on contact space in $N$-body and cosmological systems as described in \Sec\ref{sec:explaning smoothness} and \Sec\ref{ssub:conc_red_shift} respectively. We thus evade the mathematical difficulties encountered when relying on entropy to account for the AoT. This does not mean, however, that we cannot recover the notion of entropy where it does apply, as we explain in response to the next objection.
    
    
    \item {\bfseries Objections from the breakdown of thermodynamic assumptions} These objections question whether the thermodynamic assumptions underpinning Boltzmannian explanations of time-asymmetry in isolated systems of free-gases apply to self-gravitating systems in the Universe, where no preferred notion of entropy exists.

    Reassuringly, the explicit models we used to reproduce early smoothness and red-shifting towards the past do not rely on problematic assumptions about entropy being defined but rather model equilibration using dynamically monotonic functions defined explicitly within those systems. These $S$-functions, as introduced in \Sec\ref{sec:JA scenario}, behave towards attractors as one expects entropy to behave in a non-equilibrium thermodynamic system. This approach is consistent with continuing to apply thermodynamic concepts such as entropy to stars and other stellar systems such as galaxies and galaxy clusters. Thus, our approach evades the objections from the breakdown of thermodynamic assumptions by replacing the notion of Boltzmann entropy with more rigorously defined functions precisely when use of the Boltzmann entropy is questionable.
    \item {\bfseries Objections from lack of explanatory force} Sceptics worry that ``particularist'' proposals do little more than trade one brute fact (the observed arrow of time) for another (a low-entropy initial state). If the posited fact has no deeper rationale, the account seems explanatorily hollow.
    
    The JA-based framework resists this charge because its central posits --- Janus points and attractors --- are not free parameters introduced in the theory by hand but calculable consequences of taking dynamical similarity to be a gauge symmetry of the underlying equations. Once dynamical similarity is in place, natural physical assumptions are usually sufficient to imply that an entropy-minimising Janus point emerges automatically. The appeal is to a structural feature of the dynamics rather than to an unexplained contingent condition.
    
    Concrete models then reinforce the explanation with independently motivated physical inputs. In FLRW cosmology, a positive cosmological constant and matter that obeys the Weak Energy Condition guarantee late-time attractors; together with dynamical similarity they account for the observed arrow of time encoded in cosmic red-shift. In Newtonian $N$-body gravity, imposing non-negative total energy --- the analogue of a positive cosmological constant --- yields smooth configurations at the Janus point and clumped states at late times for a dense set of solutions. Because these auxiliary assumptions are backed by observation and standard theory, the resulting JA-based explanation avoids the objection that it rests on an unexplained particular fact, offering more genuine explanatory purchase than a bare Past Hypothesis.
\end{enumerate}





\section{Summary and conclusion}

In this paper, we have proposed a novel approach to accounting for the AoT. Unlike earlier approaches, it attributes the AoT neither to time-asymmetric laws nor to highly atypical initial cosmic boundary conditions. Instead, it relies first on identifying an often overlooked type of symmetry, dynamical similarity, as a gauge symmetry in cosmological theories and then on quotienting by its action in these theories. As a consequence, the quantities that have no empirical counterpart, e.g. the cosmic scale factor, disappear from the formulation of the laws.

Once dynamical similarity has been treated properly as a gauge symmetry, we find that its dynamical models often exhibit a so-called \emph{Janus-attractor} structure. Attractors are regions on which solutions focus and Janus surfaces are regions of measure preservation. JA-scenarios, we argued, exhibit a natural candidate for an entropy function, which grows towards the attractor states and has its minimum value at the Janus surface. We further showed that two key ingredients of the AoT, namely early smoothness and decreasing Hubble parameter, can, respectively, be recovered in an $N$-body and FLRW model, which exhibit a JA structure. Results from the FLRW model, in particular, are likely to generalise to more realistic models due to the cosmic no-hair conjecture. Together, these observations suggest that a JA-scenario may hold the key to a solution to the AoT problem that avoids both the problems that plague generalist and particularist accounts.

One possible objection to the JA-based account of the AoT is that theories formulated as contact systems lack theoretical virtues, such as simplicity and universality, when compared to theories formulated as symplectic systems due the appearance of the drag in the equations of motion and the failure of Liouville's theorem. The appropriate response to this objection, in our view, is that the while the theoretical virtues displayed in symplectic systems correlate highly with their usefulness and ease of handling, they come at the cost of additional representational baggage. Even if one starts with target phenomena adequately described by a contact system, one might still be tempted to ``symplectify'' the contact system by introducing an auxiliary variable that acts to artificially compensate for drag effects or measure focusing. Thus, the apparent simplicity and universality of symplectic systems can be due solely to the introduction of auxiliary structure rather than a genuine description of the phenomena.

The present work can be extended in a number of ways. It is important, for instance, to explore whether more realistic models can simultaneously explain both the smoothness and red-shift problems within a single model. Furthermore, more could be done to explain how smoothness and rapid cooling in the early Universe or the JA framework in general could be used to account for other aspects of the AoT such as the records asymmetry or the electromagnetic arrow. Finally, a quantum mechanical treatment of dynamical similarity is required to judge the applicability of our conclusions to the early Universe, where quantum fluctuations are known to have seeded large-scale structure.

The JA-based account of the AoT offers a novel and fruitful approach to one of the great problems in the foundations of physics --- one that has challenged physicists and philosophers alike since the advent of statistical mechanics.

\section*{Acknowledgements}

We would like to thank Jan--Willem Romeijn for many helpful discussions and guidance, David Sloan for valuable input and comments on a draft, and Julian Barbour for helpful remarks. SG would also like to thank Pooya Farokhi, Pedro Naranjo and Tim Koslowski for helpful comments and discussions. SG's work was partially funded by and Young Academy Groningen scholarship. SF's work was supported by the Netherlands Organization for Scientific Research (NWO), project VI.Vidi.211.088.

\bibliographystyle{apacite}
\bibliography{dissertation}

\end{document}